\documentclass[12pt]{article}
\usepackage{amsmath}
\usepackage{graphicx}
\usepackage{enumerate}
\usepackage{natbib}
\usepackage{url} %

\newcommand{\blind}{1}

\usepackage{mathrsfs}
\usepackage{amssymb}
\usepackage{amsmath}
\usepackage{ascmac}
\usepackage{amsthm}
\usepackage{natbib}
\usepackage{setspace}
\usepackage{color}
\usepackage{url}

\usepackage{booktabs}
\usepackage{multirow}
\usepackage{here}

\usepackage{comment}
\usepackage{bm}
\usepackage{bbm}

\usepackage{titlesec}
\titleformat*{\section}{\large\bfseries}
\titleformat*{\subsection}{\it}
\titleformat*{\subsubsection}{\it}

\newtheorem{thm}{Theorem}
\newtheorem{lem}{Lemma}

\newtheorem{remark}{Remark}
\def\ep{{\varepsilon}}

\def\si{{\sigma}}

\def\al{{\alpha}}
\def\be{{\beta}}
\def\ga{{\gamma}}
\def\de{{\delta}}
\def\ep{{\varepsilon}}

\def\si{{\sigma}}
\def\om{{\omega}}

\def\non{{\nonumber}}

\def\Jc{{\cal J}}
\def\Lc{{\cal L}}

\def\Kc{{\cal K}}

\def\xt{{\tilde x}}

\def\Mt{{\widetilde M}}
\def\al{{\alpha}}
\def\be{{\beta}}
\def\ga{{\gamma}}
\def\de{{\delta}}
\def\ep{{\varepsilon}}

\def\si{{\sigma}}
\def\om{{\omega}}

\def\ka{{\kappa}}

\def\bbe{{\text{\boldmath $\beta$}}}
\def\bga{{\text{\boldmath $\gamma$}}}

\def\bet{{\tilde \be}}

\def\sit{{\tilde \si}}

\def\bbet{{\widetilde \bbe}}

\def\e{{\text{\boldmath $e$}}}

\def\t{{\text{\boldmath $t$}}}

\def\x{{\text{\boldmath $x$}}}
\def\y{{\text{\boldmath $y$}}}

\def\B{{\text{\boldmath $B$}}}
\def\C{{\text{\boldmath $C$}}}
\def\D{{\text{\boldmath $D$}}}
\def\E{{\text{\boldmath $E$}}}
\def\F{{\text{\boldmath $F$}}}
\def\G{{\text{\boldmath $G$}}}

\def\I{{\text{\boldmath $I$}}}

\def\O{{\text{\boldmath $O$}}}

\def\yt{{\tilde y}}

\def\Jc{{\cal J}}
\def\Lc{{\cal L}}
\def\Kc{{\cal K}}

\def\rank{{\rm rank\,}}

\def\[{{\text{\boldmath $[$}}}
\def\]{{\text{\boldmath $]$}}}

\def\1r{{\rm (1)}}
\def\2r{{\rm (2)}}
\def\3r{{\rm (3)}}
\def\4r{{\rm (4)}}
\def\5r{{\rm (5)}}

\def\non{{\nonumber}}

\begin{document}

\def\spacingset#1{\renewcommand{\baselinestretch}%
{#1}\small\normalsize} \spacingset{1}


\if1\blind
{
  \title{\bf Posterior Robustness with Milder Conditions: Contamination Models Revisited}
  \author{
Yasuyuki Hamura\footnote{Corresponding author. 
Graduate School of Economics, Kyoto University, Yoshida-Honmachi, Sakyo-ku, Kyoto, 606-8501, JAPAN. 
\newline{
E-Mail: yasu.stat@gmail.com}}, \
Kaoru Irie\footnote{Faculty of Economics, The University of Tokyo. 
\newline{
E-Mail: irie@e.u-tokyo.ac.jp}} \
and 
Shonosuke Sugasawa\footnote{Faculty of Economics, Keio University. 
\newline{
E-Mail: sugasawa@econ.keio.ac.jp}} 
}
  \maketitle
} \fi

\if0\blind
{
  \bigskip
  \bigskip
  \bigskip
  \begin{center}
    {\LARGE\bf Posterior Robustness with Milder Conditions: Contamination Models Revisited}
\end{center}
  \medskip
} \fi

\bigskip
\begin{abstract}
Robust Bayesian linear regression is a classical but essential statistical tool. Although novel robustness properties of posterior distributions have been proved recently under a certain class of error distributions, their sufficient conditions are restrictive and exclude several important situations. 
In this work, we revisit a classical two-component mixture model for response variables, also known as contamination model, where one component is a light-tailed regression model and the other component is heavy-tailed. The latter component is independent of the regression parameters, which is crucial in proving the posterior robustness. 
We obtain new sufficient conditions for posterior (non-)robustness and reveal non-trivial robustness results by using those conditions. In particular, we find that even the Student-$t$ error distribution can achieve the posterior robustness in our framework. 
A numerical study is performed to check the Kullback-Leibler divergence between the posterior distribution based on full data and that based on data obtained by removing outliers. 
\end{abstract}

\noindent%
\vfill

\newpage
%
\section{Introduction}
\label{sec:introduction}
Bayesian posterior robustness \citep{o1979outlier} and related topics have long been studied \citep[e.g.,][]{west1984outlier, andrade2006bayesian, andrade2011bayesian, op2012}. 
There, one of the most important objectives is to perform posterior analysis using moderate observations only and discarding outliers that are not related to the parameters of interest. 
Because the task of manually detecting or determining outliers is difficult in general, robust models are desired under which the effects of outliers are automatically removed. 

Although many robust regression models have been proposed in the literature, few works \citep[e.g.,][]{o1979outlier} %
have given theoretical justifications to those models. 
In fact, it is only recently that \cite{desgagne2013, desgagne2015robustness} 
and \cite{Gag2019} %
have proved posterior robustness for scale, location-scale, and regression models, respectively. Here, posterior densities are said to be robust if they converge to the corresponding conditional densities of parameters based only on non-outliers as the absolute values of outliers tend to infinity. 
Since then, posterior robustness has been established in various practically important settings; \cite{hamura2020log} %
obtained robustness results for regressions with shrinkage priors, whereas \cite{his2021} %
considered a case of integer-valued observation.

In proving the posterior robustness, \cite{Gag2019} and \cite{hamura2020log} 
considered the following model; with observations $y_1 , \dots , y_n$, $p$-dimensional covariate vectors $\x_1,\dots ,\x_n$, regression coefficients $\bbe \in \mathbb{R} ^p$ and a scale parameter $\si \in (0, \infty )$, they assume
\begin{align} 
\label{eq:model_1} 
y_i\sim  f (( y_i - {\x _i}^{\top } \bbe ) / \si ) / \si, \ \ \ \ i=1,\ldots,n,
\end{align}
for some error density $f$ and prior $( \bbe , \si ) \sim \pi ( \bbe , \si )$. 
In their proof of posterior robustness, it is crucial to assume that $f$ is the log-regularly varying error density. 
A typical density tail of the log-regularly varying distributions is $f(y) \sim |y|^{-1} \{ \log |y| \} ^{-(\beta + 1)}$ as $|y|\to \infty$, where $\beta >0$ (For the rigorous definition, see \cite{desgagne2013}). This distribution has no finite moment and heavier density tails than those of the Student's $t$-distribution. %
If $f$ is the Student's $t$-distribution, the posterior is not robust \citep{gh2023}. These theoretical findings imply the superiority of log-regularly varying error density to the Student's $t$-distributions. However, it has also been reported that the Student's $t$-error distribution is fairly competitive in posterior inference in several numerical studies \citep{hamura2020log}.

In this paper, we revisit the following classical two-component mixture regression model, also known as the contamination model:
\begin{align}
\label{eq:model_2} 
&y_i \sim (1 - s) f_0 (( y_i - {\x _i}^{\top } \bbe ) / \si ) / \si + s f_1 ( y_i ) \text{,}  \ \ \ \ i=1,\ldots,n,
\end{align}
where $( \bbe , \si ) \sim \pi ( \bbe , \si )$ and $s \in (0, 1)$ is a prior probability that an observation becomes an outlier. 
The first density, $f_0$, has thinner tails and is typically the standard normal distribution. 
The second density, $f_1$, is a heavy-tailed distribution, such as Student's $t$-distribution, and expected to accommodate outliers. 
One notable feature of the above model is that the second term is completely independent of the parameters $( \bbe , \si )$. 
This is a significant difference from the classical two-component mixtures in \cite{box1968bayesian} and subsequent research \citep{tak2019robust,spg2020}, where the second component is also scaled by observational standard error $\sigma$. %
Scaling the second component by $\sigma$ is reasonable in terms of data fit, but could affect the inference on $\sigma$ in the presence of outliers. This observation motivates our research on the above model. %

Under the model (\ref{eq:model_2}), we show that the posterior is robust if $\pi ( \si )$, the marginal prior for $\si $, has tails sufficiently lighter than those of the error density $f_1$. 
When $f_1$ is log-regularly varying, then most of prior distributions can satisfy this sufficient condition for robustness. 
Furthermore, we prove that the sufficient condition on the tails of $\pi ( \si )$ is ``nearly'' necessary as well; if the error distribution is not log-regularly varying and has lighter tails than $\pi ( \si )$, then the posterior is not robust. %
With these conditions, we can identify the posterior (non)-robustness for most of the error and prior distributions used in the regression models. 

Our result can also explain the gap between the non-robustness of the Student $t$-distribution in model (\ref{eq:model_1}) and its success in posterior inference in numerical studies. 
For simplicity, assume that only the first observation, $y_1$, is outlying and let $|y_1|\to \infty$. 
Then, under the model (\ref{eq:model_1}) with $f(y) \propto |y|^{- 1 - \al }$ as $|y| \to \infty $ for $\al > 0$ (Student's $t$-distribution with $\al$ degree-of-freedom), it holds that 
\begin{align}
p( \bbe , \si | \y ) &\propto \pi ( \bbe , \si ) {f(( y_1 - {\x _1}^{\top } \bbe ) / \si ) \over f( y_1 ) \si } \prod_{i = 2}^{n} {f(( y_i - {\x _i}^{\top } \bbe ) / \si ) \over \si } \non \\
&\to \pi ( \bbe , \si ) \si ^{\al } \prod_{i = 2}^{n} {f(( y_i - {\x _i}^{\top } \bbe ) / \si ) \over \si } \non 
\end{align}
as $| y_1 | \to \infty $. This limit is the product of the posterior density without $y_1$ and factor $\si^{\al}$. In other words, the Student's $t$-distribution can never achieve the posterior robustness. %
By contrast, under the model (\ref{eq:model_2}), we have 
\begin{align}
p( \bbe , \si | \y ) &\propto \pi ( \bbe , \si ) \Big\{ {1 - s \over s} {f_0 (( y_1 - {\x _1}^{\top } \bbe ) / \si ) \over \si f_1 ( y_1 )} + 1 \Big\} \prod_{i = 2}^{n} \Big\{ (1 - s) {f_0 (( y_i - {\x _i}^{\top } \bbe ) / \si ) \over \si } + s f_1 ( y_i ) \Big\} \non \\
&\to \pi ( \bbe , \si ) \prod_{i = 2}^{n} \Big\{ (1 - s) {f_0 (( y_i - {\x _i}^{\top } \bbe ) / \si ) \over \si } + s f_1 ( y_i ) \Big\} \non 
\end{align}
as $| y_1 | \to \infty $, provided that $f_1$ has sufficiently heavier tails than $f_0$ %
(For the rigorous proof, including the computation of the ignored normalizing constant, see the proof of Theorem~1 in the Supplementary Materials). %
This is precisely the posterior without $y_1$, for which we confirm the posterior robustness. 
Also, note that $f_1$ can be the Student's $t$-distribution but still can achieve the posterior robustness under this model. %
The main difference from the model (\ref{eq:model_1}) is that the second component $f_1$ of (\ref{eq:model_2}) does not involve the parameters $(\bbe, \sigma)$. %
Thanks to this difference, outliers are not linked to the parameters in this model and therefore have no effects on the posterior distribution of $(\bbe , \si)$, as long as $f_1$ has heavier tails than $f_0$. 
This observation applies to the general case of multiple outliers, as will be seen below.

The remainder of this paper is organized as follows. 
In Section \ref{sec:model}, sufficient conditions and necessary conditions for posterior robustness are given. 
In Section \ref{sec:example}, a numerical example is given, in which we see that the Kullback-Leibler divergence between the target and available posteriors can diverge or converge to $0$ in some cases. 
Proofs are given in the Supplementary Material.

\section{ Contamination Models and Posterior Robustness }
\label{sec:model} 
Suppose that we observe 
\begin{align}
&y_i \sim (1 - s) {\rm{N}} ( y_i | {\x _i}^{\top } \bbe , \si ^2 ) + s f_1 ( y_i ) \non 
\end{align}
for $i = 1, \dots , n$, where $\x = ( {\x _i}^{\top } )_{i = 1}^{n}$ is a set of continuous explanatory variables and where $\bbe = ( \be _k )_{k = 1}^{p} \in \mathbb{R} ^p$ and $\si \in (0, \infty )$ are parameters of interest following a prior distribution $\pi ( \bbe , \si )$. 
Here, $f_1 ( \cdot )$ is an error density, and $s \in (0, 1)$ is a prior probability that observation is generated from $f_1$. 

Following the work of \cite{desgagne2015robustness}, %
let $\Kc , \Lc \subset \{ 1, \dots , n \} $ satisfy $\Kc \cup \Lc = \{ 1, \dots , n \} $, $\Kc \cap \Lc = \emptyset $, and $\Kc , \Lc \neq \emptyset $. 
Suppose that $a_i \in \mathbb{R}$, $b_i \neq 0$, and $y_i = a_i + b_i \om $, $\om \to \infty $, for $i \in \Lc $, such that $\Lc $ represents the set of indices of outlying observations. 
We say that the posterior is robust to outliers under the above model if $p( \bbe , \si | \y ) \to p( \bbe , \si | \y _{\Kc } )$ as $\om \to \infty $, where $\y = ( y_i )_{i = 1}^{n}$, $\y _{\Kc } = ( y_i )_{i \in \Kc }$, and $\y _{\Lc } = ( y_i )_{i \in \Lc }$. 

To derive conditions for posterior robustness, we limit the class of prior distributions for $(\bbe,\si)$. 
Suppose that 
\begin{align}
\pi ( \bbe | \si ) = {\pi ( \bbe , \si ) \over \pi ( \si )} \le M \prod_{k = 1}^{p} \Big\{ {1 \over \si } {(| \be _k | / \si )^{\ka - 1} \over (1 + | \be _k | / \si )^{\ka + \nu }} \Big\} ,\label{eq:condition_be} 
\end{align}
for some $\nu > 0$, $0 < \ka \le 1$ and $M > 0$, where $\pi ( \si ) = \int_{\mathbb{R} ^p} \pi ( \bbe , \si ) d\bbe $. 
That is, the ratio of the prior density and some double-sided scaled-beta density (with spike at the origin) must be bounded uniformly by some constant. 
This condition is satisfied by most of the conditionally independent priors that are commonly used in practice. 
Examples include shrinkage priors, such as the horseshoe prior \citep{carvalho2009handling,carvalho2010horseshoe}, as well as the normal priors. 
The condition is also satisfied by some multivariate priors for dependent $\bbe$, including the multivariate normal prior. 

Likewise, we assume the error distributions, $f_1$, are bounded as 
\begin{align}
f_1 (y) &\ge {1 \over M'} {1 \over (1 + |y|)^{1 + \al }} {1 \over \{ 1 + \log (1 + |y|) \} ^{1 + \ga }} ,\label{eq:condition_error} 
\end{align}
for some $\al \ge 0$, $\ga \ge - 1$ and $M' > 0$. 
The class of distributions that satisfy this condition includes Student's $t$-distributions ($\al > 0$ and $\ga = - 1$) and  log-regularly varying distributions ($\al = 0$ and $\ga > 0$). 

The following theorem gives a sufficient condition for the posterior to be robust. 

\begin{thm}
\label{thm:condition} 
Suppose that conditions~(\ref{eq:condition_be}) and (\ref{eq:condition_error}) are satisfied for $\nu > \al $. 
Also, suppose that 
\begin{align}
E[ \si ^{| \Lc | \al + \rho } ] < \infty \label{eq:condition_1} 
\end{align}
for some $\rho > 0$. 
Then the posterior is robust to outliers under our model; that is, we have 
\begin{align}
\lim_{\om \to \infty } p( \bbe , \si | \y ) = p( \bbe , \si | \y _{\Kc } ) \non 
\end{align}
at each $( \bbe , \si ) \in \mathbb{R} ^p \times (0, \infty )$. 
\end{thm}

The moment condition for $\pi(\si)$ in (\ref{eq:condition_1}) could be a strong requirement when $\alpha>0$ and $| \Lc |$ is large. %
We will compare this condition with those in the literature later in Table~\ref{table:condition_new}. %
Next, we prove that the posterior robustness does not hold if this moment condition is not satisfied, in addition that the error density tails are not sufficiently heavily tailed. 

\begin{thm}
\label{thm:necessary} 
Let $h \colon \mathbb{R} ^p \to (0, \infty )$ be a probability density and suppose that $\pi ( \bbe | \si ) = h( \bbe / \si ) / \si ^p$. %
Let $\al > 0$ and suppose that 
\begin{align}
f_1 (y) &\le M' {1 \over (1 + |y|)^{1 + \al }} \non 
\end{align}
for all $y \in \mathbb{R}$ for some $M' > 0$. 
Suppose that 
\begin{align}
\pi ( \si ) \ge (1 / \Mt ) / \si ^{| \Lc | \al + 1 - \rho } \label{eq:condition_2} 
\end{align}
for all $\si > 1$ for some $\Mt > 0$ and $0 < \rho < 1$. 
Then we have 
\begin{align}
\lim_{\om \to \infty } p( \bbe , \si | \y ) = 0 \non 
\end{align}
at each $( \bbe , \si ) \in \mathbb{R} ^p \times (0, \infty )$. 
\end{thm}

Clearly, under the assumptions of Theorem \ref{thm:necessary}, the posterior does not converge in the usual sense. 
Indeed, we see in the next section that the Kullback-Leibler divergence between $p( \bbe , \si | \y _{\Kc } )$ and $p( \bbe , \si | \y )$ diverges in such a situation. 

From Theorems \ref{thm:condition} and \ref{thm:necessary}, we can determine whether a prior $\pi ( \si )$ yields a robust posterior or not in most cases. 
Suppose that $\bbe / \si $ and $\si $ are independent %
(e.g., $\bbe | \sigma \sim N(0,\sigma^2)$ and $\sigma \sim \pi(\sigma)$) %
and that (\ref{eq:condition_be}) holds. 
Suppose that equality holds in (\ref{eq:condition_error}). %
Then, if we use a gamma prior for $\si ^2$, the moment condition in (\ref{eq:condition_1}) is always satisfied; hence the posterior is robust regardless of the choice of $\al $. 
If we use an inverse gamma prior or a scaled beta prior for $\si ^2$, either (\ref{eq:condition_1}) or (\ref{eq:condition_2}) is satisfied, depending on the hyperparameters. That is, there exists a threshold separating robust and non-robust cases. 
These observations are summarized in Table \ref{table:prior_new}.

\begin{table}[!htbp]
	\begin{center}
		\caption{Priors and conditions in Theorems~\ref{thm:condition} and \ref{thm:necessary}} \label{table:prior_new} 

        \vspace{6pt}
		\begin{tabular}{cccc}\toprule 
        \multirow{2}{*}{Prior for $\sigma^2$} &  \multirow{2}{*}{Density $\pi(\si)d\si$} & Condition (\ref{eq:condition_1}) & Condition (\ref{eq:condition_2}) \\
        & & for robustness & for non-robustness \\ \midrule 
        Inverse-gamma: & \multirow{2}{*}{$(1 / \si ^{2 A + 1} ) \exp (- B / \si ^2 )$} & \multirow{2}{*}{$2 A > | \Lc | \al$} & \multirow{2}{*}{$2 A < | \Lc | \al$} \\
         $\mathrm{IG}(A,B)$ & & & \\
        \multirow{2}{*}{Gamma: $\mathrm{Ga}(C,D)$} & \multirow{2}{*}{$\si ^{2 C - 1}\exp (- D \si ^2 )$} & \multirow{2}{*}{\checkmark} & \multirow{2}{*}{NA} \\
          & & & \\
        Scaled-beta: & \multirow{2}{*}{$\si ^{2 E - 1} / (1 + \si ^2 )^{E + F}$} & \multirow{2}{*}{$2 F > | \Lc | \al$} & \multirow{2}{*}{$2 F < | \Lc | \al$} \\ 
         $\mathrm{SB}(E,F)$ & & & \\ \bottomrule 
        \end{tabular}
        \end{center}
\end{table}

The sufficient conditions obtained in this study differ from those in \cite{Gag2019} and \cite{hamura2020log} not only in the model specification given in (\ref{eq:model_1}) and (\ref{eq:model_2}) but also in the requirement of the error and prior densities. Table~\ref{table:condition_new} summarizes the sufficient conditions for posterior robustness in the literature and Theorem~\ref{thm:condition}. As pointed out in the introduction, $f_1$ in our model does not have to be log-regular varying to achieve the posterior robustness, which is significantly different from the settings in the literature. Instead, at the cost of allowing for a wider class of error distributions for $f_1$, more constraints on the choice of priors for $\sigma$ are needed for the proof of Theorem~\ref{thm:condition}. 
Consequently, the conditions used in the literature and Theorem~\ref{thm:condition} are not nested in one another. For example, the conditions in \cite{Gag2019} cover the improper prior for $(\bbe,\si)$. 

It is also worth emphasizing that, as clarified in Table~\ref{table:condition_new}, no assumption is made directly on $| \Lc |$, the number of outliers, in Theorem~\ref{thm:condition}. 
Note that this number is defined by the residuals; $| y_i - {\x _i}^{\top } \bbe |$ is outlying for $i \in \Lc $ and close to zero for $i \in \Kc $. 
The key result that enables the proof without any assumption on $|\Lc|$ is the lemma we obtained about the residuals; for details, see Lemma~1 in the Supplemetary Materials.

{\small 
\begin{table}[!htbp]
	\begin{center}
         \caption{%
         Sufficient conditions of model components for robustness} \label{table:condition_new} %
        \begin{tabular}{cccccc}\toprule
         & Number of &  Error density & \multicolumn{3}{c}{ Prior density $\pi ( \bbe , \si )$}  \\
         & outliers $| \Lc |$ & tails ($f$ or $f_1$) & Density bounds & Moments & Improper \\ \midrule 
         Gagnon et al. & \multirow{2}{*}{$| \Kc | \ge | \Lc | + 2 p - 1$} & \multirow{2}{*}{LRVD} & \multirow{2}{*}{$\max \{ 1, 1 / \si \}$} & \multirow{2}{*}{--} & \multirow{2}{*}{\checkmark} \\
         (2019) & & & & \\
         Hamura et al. & \multirow{2}{*}{$| \Kc | \ge | \Lc | + p$} & \multirow{2}{*}{LRVD} & \multirow{2}{*}{$\displaystyle \sup_{t \in \mathbb{R}}  |t| \pi _{\be } (t) < \infty $} & \multirow{2}{*}{$E[ \si ^{- n} ] < \infty $} & \multirow{2}{*}{NA}  \\
         (2022) & & & & \\
         Theorem~\ref{thm:condition} & \multirow{2}{*}{Not needed} & \multirow{2}{*}{${1 \over (1 + |y|)^{1 + \al }}$} & \multirow{2}{*}{$ \prod\limits_{k = 1}^{p} {(| \be _k | / \si )^{\ka - 1} / \si \over (1 + | \be _k | / \si )^{\ka + \nu }}$} & \multirow{2}{*}{$E[ \si ^{| \Lc | \al + \rho } ] < \infty $} & \multirow{2}{*}{NA}  \\
         of this study & & & & \\
          \bottomrule 
        \end{tabular}
        \end{center}
\end{table}
}

\section{%
Numerical Examples}
\label{sec:example} 
Here, we consider a numerical example to illustrate the property of the posterior (non)-robustness. 
In doing so, we numerically evaluated the Kullback-Leibler (KL) divergence of the target posterior distribution $p( \bbe , \si | \y _{\Kc } )$ from the available posterior distribution $p( \bbe , \si | \y )$, or ${\rm{KL}} = \int_{\mathbb{R} ^p \times (0, \infty )^p} p( \bbe , \si | \y _{\Kc } ) [ \log \{ p( \bbe , \si | \y _{\Kc } ) / p( \bbe , \si | \y ) \} ] d( \bbe , \si )$, as well as the point estimates of parameters and predictive intervals. 
We used the conjugate normal-inverse gamma prior %
$\pi ( \bbe , \si ) \propto (1 / \si ^{2 A + 1} ) \exp (- B / \si ^2 ) \times (1 / \si ^p ) \exp \{ - \| \bbe \| ^2 / (2 C^2 \si ^2) \} $, 
where $A, B, C > 0$. 
Under this prior, the posterior becomes a finite mixture of known distributions and analytically and numerically tractable. 
We considered the following two error densities: 
\begin{align}
&f_{1}^{\rm{light}} (y) = {\al / 2 \over (1 + |y|)^{1 + \al }} \text{,} \quad y \in \mathbb{R} \text{,} \quad \text{and} \quad f_{1}^{\rm{heavy}} (y) = {\ga / 2 \over 1 + |y|} {1 \over \{ 1 + \log (1 + |y|) \} ^{1 + \ga }} \text{,} \quad y \in \mathbb{R} \text{,} \non 
\end{align}
where $\al , \ga > 0$. 
The first error distribution, $f_{1}^{\rm{light}}$, is the double-sided scale-beta distribution, whose tail behavior is equivalent to that of Student's $t$-distribution. The second error distribution, $f_{1}^{\rm{heavy}}$, is the unfolded version of the log-Pareto distribution of \cite{cormann2009generalizing}. %

As an example, we deterministically created the dataset as 
\begin{align}
\begin{pmatrix} \x _1 & \cdots & \x _5 \end{pmatrix} &= \begin{pmatrix} 1 & 1 & 1 & 1 & 1 \\ 1 & 2 & 3 & 4 & 5 \end{pmatrix}, \ \ \ \ \ \  \boldsymbol{\epsilon} ^{\top } = \begin{pmatrix}
    0.1 & -0.1 & 0.1 & -0.1 & \omega 
\end{pmatrix}, \ \ \ \ \ \ \boldsymbol{\beta}^{\top} = \begin{pmatrix}
    1 & 1
\end{pmatrix} \text{,} \non 
\end{align}
and $\y = \x ^{\top} \boldsymbol{\beta} + \boldsymbol{\epsilon}$. In this example, where $n = 5$ and $p = 2$, we considered $\omega \in \{ 10^{-1}, 10^{0}, 10^{1}, 10^{2}, 10^{3} \} $. 
In computing the KL divergence, the fifth observation with $\omega$ is viewed as an outlier; $\Kc = \{ 1, \dots , 4 \} $ and $\Lc = \{ 5 \} $. Our experiment includes the case of $\omega = 0.1$ to see the performance of the robust model in the absence of outliers. 
For the prior, we set $\al = 3$, $\ga = 3 / 2$, $B = C = 1$ and $s = 1 / 10$, and we considered the two cases $A = 1 / 10$ and $A = 2$. Combining the two priors with the two error distributions $f_{1}^{\rm{light}}$ and $f_{1}^{\rm{heavy}}$, we have four models in total. 

\begin{figure}[H]%
\centering
\includegraphics[width=17cm]{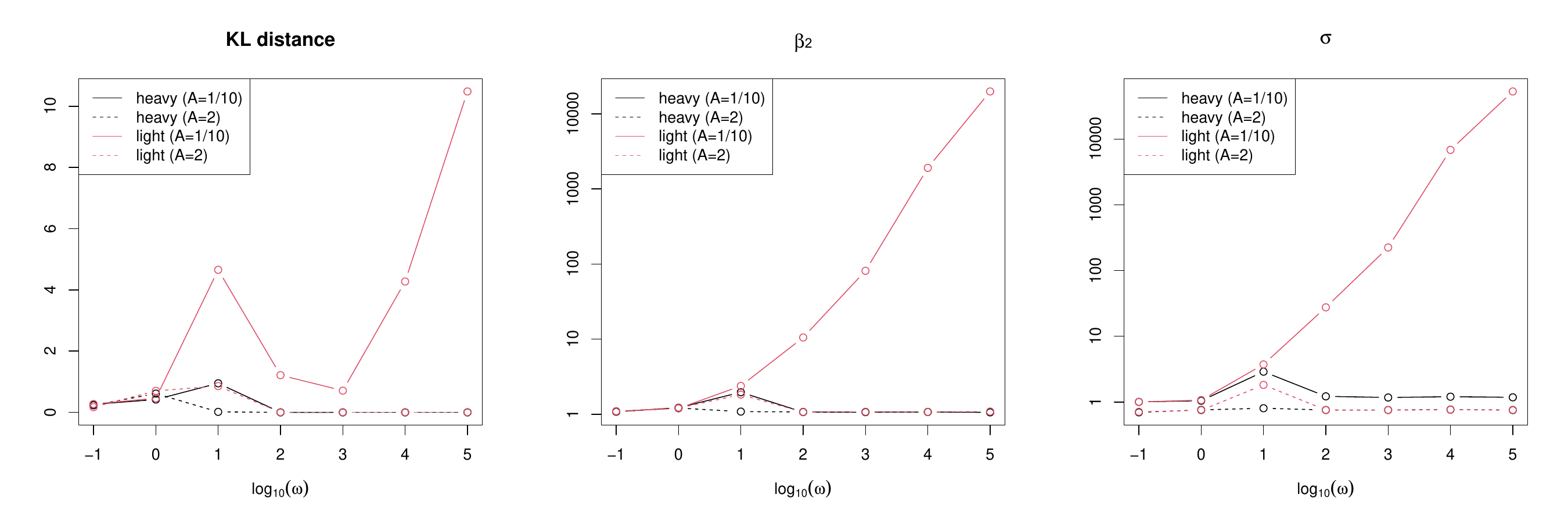
}
\caption{The KL divergence between $p( \bbe , \si | \y _{\Kc } )$ and $p( \bbe , \si | \y )$ (left), posterior means of $\be _2$ %
(center) and $\sigma$ (right) under $\om =10^{-1}, 10^0,10^1,\ldots, 10^5$. }
\label{fig:KL} 
\end{figure}

First, we obtained the Monte Carlo approximation of the KL divergence by using 1,000 samples from the posterior distributions. 
The result is summarized in the left panel of Figure~\ref{fig:KL}. 
It is clearly seen that the KL divergence does not decrease when $f_1 = f_{1}^{\rm{light}}$ and $A = 1 / 10$, 
since the condition of Theorem \ref{thm:necessary} is satisfied and the posterior is not convergent. 
In the other three cases, where the sufficient condition of Theorem \ref{thm:condition} is satisfied, the KL divergence converges to $0$ as $\omega \to \infty$. 

In addition, we computed the posterior means of $\be _2$ %
and $\sigma$ in each scenario, which are shown in the middle and right panels of Figure~\ref{fig:KL}, respectively. The point estimates of $\be _2$ %
and $\sigma$ are stable regardless of the value of $\omega$ in the three cases where the posterior robustness holds. It should also be noted that the difference of the point estimates with and without outliers (say, $\omega \ge 10^2$ and $\omega = 10^{-1}$) is small under the posterior robustness. In contrast, the point estimates become unreasonable as $\omega$ increases when $f_1 = f_{1}^{\rm{light}}$ and $A = 1 / 10$. 

Next, under the same setting, we computed the posterior and predictive distributions of $\be _1 + \xt _2 \be _2$ and $\yt \sim (1 - s) f_0 ( \{ \yt - ( \be _1 + \xt _2 \be _2 ) \} / \si ) / \si + s f_1 ( \yt )$ given $\y $ with $\om \in \{ 10^{-1}, 10^2 \}$ %
When $f_1 = f_{1}^{\rm{light}}$ and $A = 1 / 10$, the credible intervals become extremely wide since the posterior robustness does not hold and the posterior of $(\bbe,\si)$ converges to zero. 
When $A=2$ and the posterior robustness holds even for $f_{1}^{\rm{light}}$, the lengths of the interval estimates become reasonable. The interval lengths obtained under $f_{1}^{\rm{light}}$ and $f_{1}^{\rm{heavy}}$ are similar but slightly different, reflecting the difference between their error density tails.

\begin{table}[!htbp]
\caption{The lengths of $95\%$ credible and prediction intervals of $\beta_1+\tilde{x}\beta_2$ and $\tilde{y}$, respectively, evaluated at $\tilde{x}=2$ and $4$.
Here we consider $\omega=10^{-1}$ and $10^2$. 
\label{tab:pred}}
\begin{center}
\begin{tabular}{ccccccccccc}
\hline
&& \multicolumn{4}{c}{$\beta_1+\tilde{x}\beta_2$ (regression value at $\tilde{x}$)} && \multicolumn{4}{c}{$\tilde{y}$ (unobserved data at $\tilde{x}$)}\\
$(\omega, \tilde{x})$ &  & $(10^{-1}, 2)$ & $(10^{-1}, 4)$ & $(10^{2}, 2)$ & $(10^{2}, 4)$&  & $(10^{-1}, 2)$ & $(10^{-1}, 4)$ & $(10^{2}, 2)$ & $(10^{2}, 4)$\\
\hline
\hspace{-0.35cm} heavy ($A=1/10$) &  & 1.87 & 2.27 & 2.35 & 3.97 &  & 5.91 & 8.00 & 6.96 & 9.21 \\
\hspace{-0.35cm} heavy ($A=2$) &  & 1.32 & 1.62 & 1.35 & 2.37 &  & 4.63 & 6.84 & 5.02 & 7.09 \\
\hspace{-0.35cm} light ($A=1/10$) &  & 1.90 & 2.27 & 60.7 & 82.8 &  & 5.13 & 7.42 & 145 & 157 \\
\hspace{-0.35cm} light ($A=2$) &  & 1.24 & 1.56 & 1.44 & 2.39 &  & 4.51 & 6.78 & 4.48 & 6.91 \\
\hline
\end{tabular}
\end{center}
\end{table}

\section*{Acknowledgments}
Research of the authors was supported in part by JSPS KAKENHI Grant Number 22K20132, 19K11852, 17K17659, and 21H00699 from Japan Society for the Promotion of Science.

\bibliographystyle{chicago}

\bibliography{ref}

\begin{thebibliography}{}

\bibitem[\protect\citeauthoryear{Andrade and O'Hagan}{Andrade and O'Hagan}{2006}]{andrade2006bayesian}
Andrade, J. A.~A. and A.~O'Hagan (2006).
\newblock Bayesian robustness modeling using regularly varying distributions.
\newblock {\em Bayesian Analysis\/}~{\em 1\/}(1), 169--188.

\bibitem[\protect\citeauthoryear{Andrade and O'Hagan}{Andrade and O'Hagan}{2011}]{andrade2011bayesian}
Andrade, J. A.~A. and A.~O'Hagan (2011).
\newblock Bayesian robustness modelling of location and scale parameters.
\newblock {\em Scandinavian Journal of Statistics\/}~{\em 38\/}(4), 691--711.

\bibitem[\protect\citeauthoryear{Box and Tiao}{Box and Tiao}{1968}]{box1968bayesian}
Box, G.~E. and G.~C. Tiao (1968).
\newblock A bayesian approach to some outlier problems.
\newblock {\em Biometrika\/}~{\em 55\/}(1), 119--129.

\bibitem[\protect\citeauthoryear{Carvalho, Polson, and Scott}{Carvalho et~al.}{2009}]{carvalho2009handling}
Carvalho, C.~M., N.~G. Polson, and J.~G. Scott (2009).
\newblock Handling sparsity via the horseshoe.
\newblock In {\em AISTATS}, Volume~5, pp.\  73--80.

\bibitem[\protect\citeauthoryear{Carvalho, Polson, and Scott}{Carvalho et~al.}{2010}]{carvalho2010horseshoe}
Carvalho, C.~M., N.~G. Polson, and J.~G. Scott (2010).
\newblock The horseshoe estimator for sparse signals.
\newblock {\em Biometrika\/}~{\em 97\/}(2), 465--480.

\bibitem[\protect\citeauthoryear{Cormann and Reiss}{Cormann and Reiss}{2009}]{cormann2009generalizing}
Cormann, U. and R.-D. Reiss (2009).
\newblock Generalizing the pareto to the log-pareto model and statistical inference.
\newblock {\em Extremes\/}~{\em 12\/}(1), 93--105.

\bibitem[\protect\citeauthoryear{Desgagn{\'e}}{Desgagn{\'e}}{2013}]{desgagne2013}
Desgagn{\'e}, A. (2013).
\newblock Full robustness in bayesian modelling of a scale parameter.
\newblock {\em Bayesian Analysis\/}~{\em 8}, 187--220.

\bibitem[\protect\citeauthoryear{Desgagn{\'e}}{Desgagn{\'e}}{2015}]{desgagne2015robustness}
Desgagn{\'e}, A. (2015).
\newblock Robustness to outliers in location--scale parameter model using log-regularly varying distributions.
\newblock {\em The Annals of Statistics\/}~{\em 43\/}(4), 1568--1595.

\bibitem[\protect\citeauthoryear{Gagnon, Desgagne, and Bedard}{Gagnon et~al.}{2019}]{Gag2019}
Gagnon, P., P.~Desgagne, and M.~Bedard (2019).
\newblock A new bayesian approach to robustness against outliers in linear regression.
\newblock {\em Bayesian Analysis\/}~{\em 15\/}(2), 389--414.

\bibitem[\protect\citeauthoryear{Gagnon and Hayashi}{Gagnon and Hayashi}{2023}]{gh2023}
Gagnon, P. and Y.~Hayashi (2023).
\newblock Theoretical properties of bayesian student-$t$ linear regression.
\newblock {\em Statistics and Probability Letters\/}~{\em 193}.

\bibitem[\protect\citeauthoryear{Hamura, Irie, and Sugasawa}{Hamura et~al.}{2021}]{his2021}
Hamura, Y., K.~Irie, and S.~Sugasawa (2021).
\newblock Robust hierarchical modeling of counts under zero-inflation and outliers.
\newblock {\em arXiv preprint arXiv:2106.10503\/}.

\bibitem[\protect\citeauthoryear{Hamura, Irie, and Sugasawa}{Hamura et~al.}{2022}]{hamura2020log}
Hamura, Y., K.~Irie, and S.~Sugasawa (2022).
\newblock Log-regularly varying scale mixture of normals for robust regression.
\newblock {\em Computational Statistics \& Data Analysis\/}~{\em 173}, 107517.

\bibitem[\protect\citeauthoryear{O'Hagan}{O'Hagan}{1979}]{o1979outlier}
O'Hagan, A. (1979).
\newblock On outlier rejection phenomena in bayes inference.
\newblock {\em Journal of the Royal Statistical Society: Series B\/}~{\em 41\/}(3), 358--367.

\bibitem[\protect\citeauthoryear{O'Hagan and Pericchi}{O'Hagan and Pericchi}{2012}]{op2012}
O'Hagan, A. and L.~Pericchi (2012).
\newblock Bayesian heavy-tailed models and conflict resolution: A review.
\newblock {\em Brazilian Journal of Probability and Statistics\/}~{\em 26}, 372--401.

\bibitem[\protect\citeauthoryear{Silva, Prates, and Gonccalves}{Silva et~al.}{2020}]{spg2020}
Silva, N., M.~Prates, and F.~Gonccalves (2020).
\newblock Bayesian linear regression models with flexible error distributions.
\newblock {\em Journal of Statistical Computation and Simulation\/}~{\em 90}, 2571--2591.

\bibitem[\protect\citeauthoryear{Tak, Ellis, and Ghosh}{Tak et~al.}{2019}]{tak2019robust}
Tak, H., J.~A. Ellis, and S.~K. Ghosh (2019).
\newblock Robust and accurate inference via a mixture of gaussian and student’st errors.
\newblock {\em Journal of Computational and Graphical Statistics\/}~{\em 28\/}(2), 415--426.

\bibitem[\protect\citeauthoryear{West}{West}{1984}]{west1984outlier}
West, M. (1984).
\newblock Outlier models and prior distributions in bayesian linear regression.
\newblock {\em Journal of the Royal Statistical Society: Series B (Methodological)\/}~{\em 46\/}(3), 431--439.

\end{thebibliography}

\newpage
\setcounter{page}{1}
\setcounter{equation}{0}
\renewcommand{\theequation}{S\arabic{equation}}
\setcounter{section}{0}
\renewcommand{\thesection}{S\arabic{section}}
\setcounter{table}{0}
\renewcommand{\thetable}{S\arabic{table}}
\setcounter{figure}{0}
\renewcommand{\thefigure}{S\arabic{figure}}

\begin{center}
{\LARGE\bf Supplementary Material for ``Posterior Robustness with Milder Conditions: Contamination Models Revisited''}
\end{center}

\section{A Basic Lemma}
Lemma \ref{lem:general} is used in the proof of Theorem 1. %
If $m, \tilde{m} \in \mathbb{N}$ satisfy $\tilde{m} \le m$, we write $\e _{\tilde{m}}^{(m)}$ for the $\tilde{m}$th unit vector in $\mathbb{R} ^m$, namely the $\tilde{m}$th column of the $m \times m$ identity matrix.

\begin{lem}
\label{lem:general} 
Let $n, p \in \mathbb{N}$. 
Let $\x _1 , \dots , \x _n \in \mathbb{R} ^p$ be continuous variables. 
Let $( a_1 , b_1 ), \dots , ( a_n , b_n ) \in \mathbb{R} ^2$. 
Let $\Kc = \{ i = 1, \dots , n | b_i = 0 \} $ and $\Lc = \{ 1, \dots , n \} \setminus \Kc $. 
Let $\Jc = \{ - 1, \dots , - p \} \cup \Kc $. 
Let $a_j = 0$ and $\x _j = \e _{- j}^{(p)}$ for $j = - 1, \dots , - p$. 
Suppose that $\Lc \neq \emptyset $. 
Suppose that $\{ b_i | i \in \Lc \} $ and $\{ a_i | i \in \Kc \} $ are continuous variables.

\begin{itemize}
\item[{\rm{(i)}}]
Let $1 \le l \le | \Lc |$. 
Let $i_1 , \dots , i_l \in \Lc $ satisfy $i_1 < \dots < i_l$. 
Let $\overline{\ep } > 0$ and $\underline{\om } > 0$ be arbitrary. 
Then there exist $\eta > 0$, $\de > 0$, $0 < \ep < \overline{\ep }$, and $M > \underline{\om }$ such that for all $\om \ge M$ and all $\bbe = ( \be _k )_{k = 1}^{p} \in \mathbb{R} ^p$, the condition that 
\begin{align}
| a_{i_1} + b_{i_1} \om - {\x _{i_1}}^{\top } \bbe |, \dots , | a_{i_l} + b_{i_l} \om - {\x _{i_l}}^{\top } \bbe | \le \ep \om \non 
\end{align}
implies the following conditions: 
\begin{itemize}
\item[{\rm{(a)}}]
There exist distinct indices $k_1 , \dots , k_l = 1, \dots , p$ such that $| \be _{k_1} |, \dots , | \be _{k_l} | > \de \om $. 
\item[{\rm{(b)}}]
There exist distinct indices $j_1 , \dots , j_{| \Kc | + l} \in \Jc $ such that $| a_{j_1} - {\x _{j_1}}^{\top } \bbe |, \dots , | a_{j_{| \Kc | + l}} - {\x _{j_{| \Kc | + l}}}^{\top } \bbe | > \eta $. 
\end{itemize}
\item[{\rm{(ii)}}]
Let $\overline{\ep } > 0$ and $\underline{\om } > 0$ be arbitrary. 
Then there exist $\eta > 0$, $\de > 0$, $0 < \ep < \overline{\ep }$, and $M > \underline{\om }$ such that for all $\om \ge M$, 
\begin{align}
\mathbb{R} ^p &\subset \Big( \bigcap_{i \in \Lc } \{ \bbe \in \mathbb{R} ^p | | a_i + b_i \om - {\x _i}^{\top } \bbe | > \ep \om \} \Big) \non \\
&\quad \cup \bigcup_{l = 1}^{\min \{ | \Lc |, p \} } \bigcup_{\substack{i_1 , \dots , i_l \in \Lc \\ i_1 < \dots < i_l}} \Big( \Big( \bigcap_{i \in \{ i_1 , \dots , i_l \} } \{ \bbe \in \mathbb{R} ^p | | a_i + b_i \om - {\x _i}^{\top } \bbe | \le \ep \om \} \Big) \non \\
&\quad \cap \Big( \bigcap_{i \in \Lc \setminus \{ i_1 , \dots , i_l \} } \{ \bbe \in \mathbb{R} ^p | | a_i + b_i \om - {\x _i}^{\top } \bbe | > \ep \om \} \Big) \non \\
&\quad \cap \Big\{ \bigcup_{1 \le k_1 < \dots < k_l \le p} \bigcap_{k \in \{ k_1 , \dots , k_l \} } \{ \bbe \in \mathbb{R} ^p | |( \e _{k}^{(p)} )^{\top } \bbe | \ge \de \om \} \Big\} \non \\
&\quad \cap \Big[ \Big( \bigcap_{j \in \Jc } \{ \bbe \in \mathbb{R} ^p | | a_j - {\x _j}^{\top } \bbe | > \eta \} \Big) \non \\
&\quad \cup \bigcup_{1 \le q \le p - l} \bigcup_{\substack{j_1 , \dots , j_q \in \Jc \\ j_1 < \dots < j_q}} \Big\{ \Big( \bigcap_{j \in \{ j_1 , \dots , j_q \} } \{ \bbe \in \mathbb{R} ^p | | a_j - {\x _j}^{\top } \bbe | \le \eta \} \Big) \non \\
&\quad \cap \Big( \bigcap_{j \in \Jc \setminus \{ j_1 , \dots , j_q \} } \{ \bbe \in \mathbb{R} ^p | | a_j - {\x _j}^{\top } \bbe | > \eta \} \Big) \Big\} \Big] \Big) \text{.} \non 
\end{align}
\end{itemize}
\end{lem}

\begin{proof}%
Part (ii) follows from part (i). 
For part (i), fix $\eta , \de , \ep , M > 0$ and $\om \ge M$ and $\bbe = ( \be _k )_{k = 1}^{p} \in \mathbb{R} ^p$. 
Suppose that $| a_{i_1} + b_{i_1} \om - {\x _{i_1}}^{\top } \bbe |, \dots , | a_{i_l} + b_{i_l} \om - {\x _{i_l}}^{\top } \bbe | \le \ep \om $. 
Then 
\begin{align}
( {\x _{i_h}}^{\top } )_{h = 1}^{l} \bbe \in ( a_{i_h} )_{h = 1}^{l} + ( b_{i_h} )_{h = 1}^{l} \om + \om [ \pm \ep ]^l \text{.} \non 
\end{align}
If $M > 0$ is sufficiently large, 
\begin{align}
( {\x _{i_h}}^{\top } )_{h = 1}^{l} \bbe \in ( b_{i_h} )_{h = 1}^{l} \om + \om [ \pm 2 \ep ]^l \text{.} \label{lgeneralp1} 
\end{align}
Now, suppose that $l \ge p + 1$. 
Then 
\begin{align}
( {\x _{i_h}}^{\top } )_{h = 1}^{p + 1} \bbe / \om \in ( b_{i_h} )_{h = 1}^{p + 1} + [ \pm 2 \ep ]^{p + 1} \text{.} \non 
\end{align}
Since $( {\x _{i_h}}^{\top } )_{h = 1}^{p}$ is invertible by assumption, 
\begin{align}
{\x _{i_{p + 1}}}^{\top } (( {\x _{i_h}}^{\top } )_{h = 1}^{p} )^{- 1} (( b_{i_h} )_{h = 1}^{p} + \t ) = b_{i_{p + 1}} + t \non 
\end{align}
for some $\t \in [ \pm 2 \ep ]^p$ and $t \in [ \pm 2 \ep ]$. 
This is a contradiction if $\ep > 0$ is sufficiently small since ${\x _{i_{p + 1}}}^{\top } (( {\x _{i_h}}^{\top } )_{h = 1}^{p} )^{- 1} ( b_{i_h} )_{h = 1}^{p} \neq b_{i_{p + 1}}$ by assumption. 
Thus, we have $l \le p$ if $\ep $ is sufficiently small and we assume that $l \le p$. 

For part (a), suppose that there exist distinct indices $k_1 , \dots , k_{p - l + 1} = 1, \dots , p$ such that $| \be _{k_1} |, \dots , | \be _{k_{p - l + 1}} | \le \de \om $. 
Then 
\begin{align}
(( \e _{k_h}^{(p)} )^{\top } )_{h = 1}^{p - l + 1} \bbe \in \om [ \pm \de ]^{p - l + 1} \text{.} \label{lgeneralp2} 
\end{align}
Let $k_1' , \dots , k_{l - 1}' \in \{ 1, \dots , p \} \setminus \{ k_1 , \dots , k_{p - l + 1} \} $ be such that $k_1' < \dots < k_{l - 1}'$. 
Let $\E = (( \e _{k_h'}^{(p)} )^{\top } )_{h = 1}^{l - 1}$. 
Then if $\de > 0$ is sufficiently small, by (\ref{lgeneralp1}) and (\ref{lgeneralp2}) 
\begin{align}
( {\x _{i_h}}^{\top } \E ^{\top } )_{h = 1}^{l} \E \bbe \in ( b_{i_h} )_{h = 1}^{l} \om + \om [ \pm 3 \ep ]^l \non 
\end{align}
or 
\begin{align}
( {\x _{i_h}}^{\top } \E ^{\top } )_{h = 1}^{l} \E \bbe / \om \in ( b_{i_h} )_{h = 1}^{l} + [ \pm 3 \ep ]^l \text{.} \non 
\end{align}
Let $s = \rank ( {\x _{i_h}}^{\top } \E ^{\top } )_{h = 1}^{l} \le l - 1$. 
Then there exist invertible matrices $\F \in \mathbb{R} ^{l \times l}$ and $\G \in \mathbb{R} ^{(l - 1) \times (l - 1)}$ such that 
\begin{align}
\F ( {\x _{i_h}}^{\top } )_{h = 1}^{l} \E ^{\top } \G = \begin{pmatrix} \I ^{(s)} & \O ^{(s, l - 1 - s)} \\ \O ^{(l - s, s)} & \O ^{(l - s, l - 1 - s)} \end{pmatrix} \text{.} \non 
\end{align}
Therefore, 
\begin{align}
\F ^{- 1} \begin{pmatrix} \I ^{(s)} & \O ^{(s, l - 1 - s)} \\ \O ^{(l - s, s)} & \O ^{(l - s, l - 1 - s)} \end{pmatrix} \G ^{- 1} \E \bbe / \om \in ( b_{i_h} )_{h = 1}^{l} + [ \pm 3 \ep ]^l \text{.} \non 
\end{align}
Thus, there exists $\bga \in \mathbb{R} ^s$ such that 
\begin{align}
\F ^{- 1} \begin{pmatrix} \I ^{(s)} \\ \O ^{(l - s, s)} \end{pmatrix} \bga \in ( b_{i_h} )_{h = 1}^{l} + [ \pm 3 \ep ]^l \non 
\end{align}
or 
\begin{align}
\begin{pmatrix} \bga \\ \bm{0} ^{(l - s)} \end{pmatrix} \in \F ( b_{i_h} )_{h = 1}^{l} + \F [ \pm 3 \ep ]^l \text{,} \non 
\end{align}
which is a contradiction if $\ep > 0$ is sufficiently small since $0 \neq ( \e _{l}^{(l)} )^{\top } \F ( b_{i_h} )_{h = 1}^{l}$ by assumption. 
This proves part (a). 

For part (b), suppose that there exist distinct indices $j_1 , \dots , j_{p - l + 1} \in \Jc $ such that $| a_{j_1} - {\x _{j_1}}^{\top } \bbe |, \dots , | a_{j_{p - l + 1}} - {\x _{j_{p - l + 1}}}^{\top } \bbe | \le \eta $. 
Then 
\begin{align}
( {\x _{j_h}}^{\top } )_{h = 1}^{p - l + 1} \bbe \in ( a_{j_h} )_{h = 1}^{p - l + 1} + [ \pm \eta ]^{p - l + 1} \text{.} \non 
\end{align}
Let $q_1 , q_2 \ge 0$, $j_1' , \dots , j_{q_1}' = - 1, \dots , - p$, and $j_1'' , \dots , j_{q_2}'' \in \Kc $ be such that $q_1 + q_2 = p - l + 1$, $j_1' < \dots < j_{q_1}' < j_1'' < \dots < j_{q_2}''$, and $\{ j_1' , \dots , j_{q_1}' \} \cup \{ j_1'' , \dots , j_{q_2}'' \} = \{ j_1 , \dots , j_{p - l + 1} \} $. 
Then 
\begin{align}
( \be _{- j_h'} )_{h = 1}^{q_1} \in [ \pm \eta ]^{q_1} \text{.} \non 
\end{align}
Let $1 \le k_1 < \dots < k_{p - q_1} \le p$ be such that $\{ k_1 , \dots , k_{p - q_1} \} = \{ 1, \dots , p \} \setminus \{ - j_1' , \dots , - j_{q_1}' \} $ and let $\E = ( \e _{k_1}^{(p)} , \dots , \e _{k_{p - q_1}}^{(p)} )$. 
Then if $M > 0$ is sufficiently large, we have, by (\ref{lgeneralp1}), 
\begin{align}
( {\x _{i_h}}^{\top } \E )_{h = 1}^{l} \E ^{\top } \bbe \in ( b_{i_h} )_{h = 1}^{l} \om + \om [ \pm 3 \ep ]^l \text{.} \label{lgeneralp3} 
\end{align}
Also, 
\begin{align}
( {\x _{j_h''}}^{\top } \E )_{h = 1}^{q_2} \E ^{\top } \bbe \in ( a_{j_h''} )_{h = 1}^{q_2} + [ \pm (1 + q_1 A) \eta ]^{q_2} \text{,} \non 
\end{align}
where $A = \max_{1 \le i \le n} \max_{1 \le k \le p} | {\x _i}^{\top } \e _{k}^{(p)} |$. 
If $\eta > 0$ is sufficiently small, the matrix $( {\x _{j_h''}}^{\top } \E )_{h = 1}^{q_2}$ has rank $q_2$ since otherwise $0 \in ( \e _{q_2}^{( q_2 )} )^{\top } \C ( a_{j_h''} )_{h = 1}^{q_2} + ( \e _{q_2}^{( q_2 )} )^{\top } \C [ \pm (1 + q_1 A) \eta ]^{q_2}$ for some invertible matrix $\C \in \mathbb{R} ^{q_2 \times q_2}$. 
Therefore, there exists an invertible matrix $\D \in \mathbb{R} ^{(p - q_1 ) \times (p - q_1 )}$ such that 
\begin{align}
( \I ^{( q_2 )} ,  \O ^{( q_2 , p - q_1 - q_2 )}) \bga = ( {\x _{j_h''}}^{\top } \E )_{h = 1}^{q_2} \D \D ^{- 1} \E ^{\top } \bbe \in ( a_{j_h''} )_{h = 1}^{q_2} + [ \pm (1 + q_1 A) \eta ]^{q_2} \text{,} \label{lgeneralp4} 
\end{align}
where $\bga = ( \ga _h )_{h = 1}^{p - q_1} = \D ^{- 1} \E ^{\top } \bbe $. 
It follows from (\ref{lgeneralp3}) and (\ref{lgeneralp4}) that if $M > 0$ is sufficiently large, 
\begin{align}
( {\x _{i_h}}^{\top } \E )_{h = 1}^{l} \D ( \e _{q_2 + 1}^{(p - q_1 )} , \dots , \e _{p - q_1}^{(p - q_1 )} ) ( \ga _h )_{h = q_2 + 1}^{p - q_1} / \om \in ( b_{i_h} )_{h = 1}^{l} + [ \pm 4 \ep ]^l \text{.} \non 
\end{align}
Thus, since the rank of the matrix $( {\x _{i_h}}^{\top } \E )_{h = 1}^{l} \D ( \e _{q_2 + 1}^{(p - q_1 )} , \dots , \e _{p - q_1}^{(p - q_1 )} )$ is less than or equal to $p - q_1 - q_2 = l - 1$, 
\begin{align}
0 \in ( \e _{p - q_1}^{(p - q_1 )} )^{\top } \B ( b_{i_h} )_{h = 1}^{l} + ( \e _{p - q_1}^{(p - q_1 )} )^{\top } \B [ \pm 4 \ep ]^l \non 
\end{align}
for some invertible matrix $\B \in \mathbb{R} ^{l \times l}$, which is a contradiction if $\ep > 0$ is sufficiently small. 
This completes the proof. 
\end{proof}

\section{Proof of Theorem 1
}

Here, we prove Theorem 
1.

\bigskip

\noindent
{\bf Proof of Theorem 
1.} \ \ The posterior is 
\begin{align}
p( \bbe , \si | \y ) &= \frac{ \displaystyle g( \bbe , \si ; \om ) }{ \displaystyle \int_{\mathbb{R} ^p \times (0, \infty )} g( \bbe , \si ; \om ) d( \bbe , \si ) } \text{,} \non 
\end{align}
where 
\begin{align}
g( \bbe , \si ; \om ) %
&= \pi ( \bbe , \si ) \Big[ \prod_{i \in \Kc } \Big\{ {1 - s \over s} {{\rm{N}} ( y_i | {\x _i}^{\top } \bbe , \si ^2 ) \over f_1 ( y_i )} + 1 \Big\} \Big] \prod_{i \in \Lc } \Big\{ {1 - s \over s} {{\rm{N}} ( y_i | {\x _i}^{\top } \bbe , \si ^2 ) \over f_1 ( y_i )} + 1 \Big\} \text{.} \non 
\end{align}
Since 
\begin{align}
\lim_{\om \to \infty } \prod_{i \in \Lc } \Big\{ {1 - s \over s} {{\rm{N}} ( y_i | {\x _i}^{\top } \bbe , \si ^2 ) \over f_1 ( y_i )} + 1 \Big\} = 1 \text{,} \non 
\end{align}
it is sufficient to show that 
\begin{align}
\lim_{\om \to \infty } \int_{\mathbb{R} ^p \times (0, \infty )} g( \bbe , \si ; \om ) d( \bbe , \si ) &= \int_{\mathbb{R} ^p \times (0, \infty )} \pi ( \bbe , \si ) \Big[ \prod_{i \in \Kc } \Big\{ {1 - s \over s} {{\rm{N}} ( y_i | {\x _i}^{\top } \bbe , \si ^2 ) \over f_1 ( y_i )} + 1 \Big\} \Big] d( \bbe , \si ) \text{.} \non 
\end{align}
Since for all $\ep > 0$ and all $i \in \Lc $, $| y_i - {\x _i}^{\top } \bbe | \ge \ep \om $ and $| y_i | \ge 1$ imply 
\begin{align}
{{\rm{N}} ( y_i | {\x _i}^{\top } \bbe , \si ^2 ) \over f_1 ( y_i )} &\le {M' \over \sqrt{2 \pi }} {(1 + | y_i |)^{1 + \al } \{ 1 + \log (1 + | y_i |) \} ^{1 + \ga } \over \si \exp \{ (| y_i - {\x _i}^{\top } \bbe | / \si )^2 / 2 \} } \non \\
&\le {M' \over \sqrt{2 \pi }} 2^{1 + \al } \si ^{\al } \{ 1 + \log (1 + \si ) \} ^{1 + \ga } {(| y_i | / \si )^{1 + \al } \{ 1 + \log (1 + | y_i | / \si ) \} ^{1 + \ga } \over \exp \{ \ep ^2 ( \om / \si )^2 / 2 \} } \non \\
&\le M_1 \si ^{\al } \{ 1 + \log (1 + \si ) \} ^{1 + \ga } \non 
\end{align}
for some $M_1 > 0$, it follows from the dominated convergence theorem that 
\begin{align}
&\lim_{\om \to \infty } \int_{\mathbb{R} ^p \times (0, \infty )} \Big\{ \prod_{i \in \Lc } 1(| y_i - {\x _i}^{\top } \bbe | \ge \ep \om ) \Big\} g( \bbe , \si ; \om ) d( \bbe , \si ) \non \\
&= \int_{\mathbb{R} ^p \times (0, \infty )} \pi ( \bbe , \si ) \Big[ \prod_{i \in \Kc } \Big\{ {1 - s \over s} {{\rm{N}} ( y_i | {\x _i}^{\top } \bbe , \si ^2 ) \over f_1 ( y_i )} + 1 \Big\} \Big] d( \bbe , \si ) \non 
\end{align}
for all $0 < \ep < \min_{i \in \Lc } | b_i | / 2$. 
Thus, since 
\begin{align}
g( \bbe , \si ; \om ) %
&= \sum_{\widetilde{\Kc } \subset \Kc } \sum_{\widetilde{\Lc } \subset \Lc } \pi ( \bbe , \si ) \Big( {1 - s \over s} \Big) ^{| \widetilde{\Kc } | + | \widetilde{\Lc } |} \Big\{ \prod_{i \in \widetilde{\Kc }} {{\rm{N}} ( y_i | {\x _i}^{\top } \bbe , \si ^2 ) \over f_1 ( y_i )} \Big\} \prod_{i \in \widetilde{\Lc }} {{\rm{N}} ( y_i | {\x _i}^{\top } \bbe , \si ^2 ) \over f_1 ( y_i )} \text{,} \non 
\end{align}
it suffices to prove that for all $\widetilde{\Kc } \subset \Kc $ and all $\widetilde{\Lc } \subset \Lc $, there exists $0 < \ep < \min_{i \in \Lc } | b_i | / 2$ such that 
\begin{align}
&\lim_{\om \to \infty } \int_{\mathbb{R} ^p \times (0, \infty )} h_{\widetilde{\Kc } , \widetilde{\Lc }} ( \bbe , \si ; \om ; \ep ) d( \bbe , \si ) = 0 \text{,} \non 
\end{align}
where 
\begin{align}
h_{\widetilde{\Kc } , \widetilde{\Lc }} ( \bbe , \si ; \om ; \ep ) &= \Big\{ 1 - \prod_{i \in \Lc } 1(| y_i - {\x _i}^{\top } \bbe | \ge \ep \om ) \Big\} \pi ( \bbe , \si ) \Big\{ \prod_{i \in \widetilde{\Kc }} {{\rm{N}} ( y_i | {\x _i}^{\top } \bbe , \si ^2 ) \over f_1 ( y_i )} \Big\} \prod_{i \in \widetilde{\Lc }} {{\rm{N}} ( y_i | {\x _i}^{\top } \bbe , \si ^2 ) \over f_1 ( y_i )} \non 
\end{align}
converges to $0$ as $\om \to \infty $. 
This clearly holds for $\widetilde{\Lc } = \emptyset $ for all $\widetilde{\Kc } \subset \Kc $. 

First, fix $\emptyset \neq \widetilde{\Lc } \subset \Lc $ and let $\widetilde{\Kc } = \emptyset $. 
Then, by Lemma \ref{lem:general}, there exist $\de > 0$, $0 < \ep < \min_{i \in \Lc } | b_i | / 2$, and $M > 0$ such that for all $\om \ge M$, 
\begin{align}
\mathbb{R} ^p &\subset \Big( \bigcap_{i \in \widetilde{\Lc }} \{ \bbe \in \mathbb{R} ^p | | a_i + b_i \om - {\x _i}^{\top } \bbe | > \ep \om \} \Big) \non \\
&\quad \cup \bigcup_{l = 1}^{\min \{ | \widetilde{\Lc } |, p \} } \bigcup_{\substack{i_1 , \dots , i_l \in \widetilde{\Lc } \\ i_1 < \dots < i_l}} \Big\{ \Big( \bigcap_{i \in \{ i_1 , \dots , i_l \} } \{ \bbe \in \mathbb{R} ^p | | a_i + b_i \om - {\x _i}^{\top } \bbe | \le \ep \om \} \Big) \non \\
&\quad \cap \Big( \bigcap_{i \in \widetilde{\Lc } \setminus \{ i_1 , \dots , i_l \} } \{ \bbe \in \mathbb{R} ^p | | a_i + b_i \om - {\x _i}^{\top } \bbe | > \ep \om \} \Big) \non \\
&\quad \cap \bigcup_{1 \le k_1 < \dots < k_l \le p} \bigcap_{k \in \{ k_1 , \dots , k_l \} } \{ \bbe \in \mathbb{R} ^p | | ( \e _{k}^{(p)} )^{\top } \bbe | \ge \de \om \} \Big\} \text{.} \non 
\end{align}
Since 
\begin{align}
h_{\widetilde{\Kc } , \widetilde{\Lc }} ( \bbe , \si ; \om ; \ep ) &= \Big\{ 1 - \prod_{i \in \Lc } 1(| y_i - {\x _i}^{\top } \bbe | \ge \ep \om ) \Big\} \pi ( \bbe , \si ) \prod_{i \in \widetilde{\Lc }} {{\rm{N}} ( y_i | {\x _i}^{\top } \bbe , \si ^2 ) \over f_1 ( y_i )} \non 
\end{align}
for all $\om > 0$, clearly 
\begin{align}
\lim_{\om \to \infty } \int_{\mathbb{R} ^p \times (0, \infty )} 1 \Big( \bbe \in \bigcap_{i \in \widetilde{\Lc }} \{ \bbet \in \mathbb{R} ^p | | a_i + b_i \om - {\x _i}^{\top } \bbet | > \ep \om \} \Big) h_{\widetilde{\Kc } , \widetilde{\Lc }} ( \bbe , \si ; \om ) d( \bbe , \si ) &= 0 \non 
\end{align}
by the dominated convergence theorem. 
Fix $1 \le l \le \min \{ | \widetilde{\Lc } |, p \} $, $i_1 , \dots , i_l \in \widetilde{\Lc }$ with $i_1 < \dots < i_l$, and $1 \le k_1 < \dots < k_l \le p$ and let 
\begin{align}
A( \om ; \ep , \de ) &= \Big( \bigcap_{i \in \{ i_1 , \dots , i_l \} } \{ \bbe \in \mathbb{R} ^p | | a_i + b_i \om - {\x _i}^{\top } \bbe | \le \ep \om \} \Big) \non \\
&\quad \cap \Big( \bigcap_{i \in \widetilde{\Lc } \setminus \{ i_1 , \dots , i_l \} } \{ \bbe \in \mathbb{R} ^p | | a_i + b_i \om - {\x _i}^{\top } \bbe | > \ep \om \} \Big) \non \\
&\quad \cap \bigcap_{k \in \{ k_1 , \dots , k_l \} } \{ \bbe \in \mathbb{R} ^p | | ( \e _{k}^{(p)} )^{\top } \bbe | \ge \de \om \} \text{.} \non 
\end{align}
Then 
\begin{align}
&1( \bbe \in A( \om ; \ep , \de )) h_{\widetilde{\Kc } , \widetilde{\Lc }} ( \bbe , \si ; \om ) \non \\
&\le M 1( \bbe \in A( \om ; \ep , \de )) \pi ( \si ) \Big\{ \prod_{k = 1}^{p} {(| \be _k | / \si )^{\ka - 1} / \si \over (1 + | \be _k | / \si )^{\ka + \nu }} \Big\} \prod_{i \in \widetilde{\Lc }} {{\rm{N}} ( y_i | {\x _i}^{\top } \bbe , \si ^2 ) \over f_1 ( y_i )} \non \\
&\le M_2 1( \bbe \in A( \om ; \ep , \de )) \pi ( \si ) \Big\{ \prod_{k \in \{ 1, \dots , p \} \setminus \{ k_1 , \dots , k_l \} } {(| \be _k | / \si )^{\ka - 1} / \si \over (1 + | \be _k | / \si )^{\ka + \nu }} \Big\} \Big( \prod_{i \in \{ i_1 , \dots , i_l \} } {\rm{N}} ( y_i | {\x _i}^{\top } \bbe , \si ^2 ) \Big) \non \\
&\quad \times \Big\{ \prod_{i \in \{ i_1 , \dots , i_l \} } {1 \over f_1 ( y_i )} \Big\} \Big\{ \prod_{k \in \{ k_1 , \dots , k_l \} } {( \de \om / \si )^{\ka - 1} / \si \over (1 + \de \om / \si )^{\ka + \nu }} \Big\} [ \si ^{\al } \{ 1 + \log (1 + \si ) \} ^{1 + \ga } ]^{| \widetilde{\Lc } \setminus \{ i_1 , \dots , i_l \} |} \non \\
&\le M_3 \pi ( \si ) \Big\{ \prod_{k \in \{ 1, \dots , p \} \setminus \{ k_1 , \dots , k_l \} } {(| \be _k | / \si )^{\ka - 1} / \si \over (1 + | \be _k | / \si )^{\ka + \nu }} \Big\} \Big( \prod_{i \in \{ i_1 , \dots , i_l \} } {\rm{N}} ( y_i | {\x _i}^{\top } \bbe , \si ^2 ) \Big) \non \\
&\quad \times \Big\{ \prod_{i \in \{ i_1 , \dots , i_l \} } {\om ^{\al } ( \log \om )^{1 + \ga } \over \om ^{{\al }'}} \Big\} \si ^{l {\al }'} [ \si ^{\al } \{ 1 + \log (1 + \si ) \} ^{1 + \ga } ]^{| \widetilde{\Lc } \setminus \{ i_1 , \dots , i_l \} |} \non 
\end{align}
for some $M_2 , M_3 > 0$ for any $\al < {\al }' \le \nu $ and therefore 
\begin{align}
&\lim_{\om \to \infty } \int_{\mathbb{R} ^p \times (0, \infty )} 1( \bbe \in A( \om ; \ep , \de )) h_{\widetilde{\Kc } , \widetilde{\Lc }} ( \bbe , \si ; \om ) d( \bbe , \si ) = 0 \non 
\end{align}
for some $\al < {\al }' \le \nu $. 

Next, fix $\emptyset \neq \widetilde{\Lc } \subset \Lc $ and $\emptyset \neq \widetilde{\Kc } \subset \Kc $. 
Let $\widetilde{\Jc } = \{ - 1, \dots , - p \} \cup \widetilde{\Kc }$. 
Let $y_j = 0$ and $\x _j = \e _{- j}^{(p)}$ for $j = - 1, \dots , - p$. 
Then, by Lemma \ref{lem:general}, there exist $\eta > 0$, $\de > 0$, $0 < \ep < \min_{i \in \Lc } | b_i | / 2$, and $M > 0$ such that for all $\om \ge M$, 
\begin{align}
\mathbb{R} ^p &\subset \Big( \bigcap_{i \in \widetilde{\Lc }} \{ \bbe \in \mathbb{R} ^p | | a_i + b_i \om - {\x _i}^{\top } \bbe | > \ep \om \} \Big) \non \\
&\quad \cup \bigcup_{l = 1}^{\min \{ | \widetilde{\Lc } |, p \} } \bigcup_{\substack{i_1 , \dots , i_l \in \widetilde{\Lc } \\ i_1 < \dots < i_l}} \Big( \Big( \bigcap_{i \in \{ i_1 , \dots , i_l \} } \{ \bbe \in \mathbb{R} ^p | | a_i + b_i \om - {\x _i}^{\top } \bbe | \le \ep \om \} \Big) \non \\
&\quad \cap \Big( \bigcap_{i \in \widetilde{\Lc } \setminus \{ i_1 , \dots , i_l \} } \{ \bbe \in \mathbb{R} ^p | | a_i + b_i \om - {\x _i}^{\top } \bbe | > \ep \om \} \Big) \non \\
&\quad \cap \Big\{ \bigcup_{1 \le k_1 < \dots < k_l \le p} \bigcap_{k \in \{ k_1 , \dots , k_l \} } \{ \bbe \in \mathbb{R} ^p | |( \e _{k}^{(p)} )^{\top } \bbe | \ge \de \om \} \Big\} \non \\
&\quad \cap \Big[ \Big( \bigcap_{j \in \widetilde{\Jc }} \{ \bbe \in \mathbb{R} ^p | | y_j - {\x _j}^{\top } \bbe | > \eta \} \Big) \non \\
&\quad \cup \bigcup_{1 \le q \le p - l} \bigcup_{\substack{j_1 , \dots , j_q \in \widetilde{\Jc } \\ j_1 < \dots < j_q}} \Big\{ \Big( \bigcap_{j \in \{ j_1 , \dots , j_q \} } \{ \bbe \in \mathbb{R} ^p | | y_j - {\x _j}^{\top } \bbe | \le \eta \} \Big) \non \\
&\quad \cap \Big( \bigcap_{j \in \widetilde{\Jc } \setminus \{ j_1 , \dots , j_q \} } \{ \bbe \in \mathbb{R} ^p | | y_j - {\x _j}^{\top } \bbe | > \eta \} \Big) \Big\} \Big] \Big) \text{.} \non 
\end{align}
Clearly, 
\begin{align}
&\lim_{\om \to \infty } \int_{\mathbb{R} ^p \times (0, \infty )} 1\Big( \bbe \in \bigcap_{i \in \widetilde{\Lc }} \{ \bbet \in \mathbb{R} ^p | | a_i + b_i \om - {\x _i}^{\top } \bbet | > \ep \om \} \Big) h_{\widetilde{\Kc } , \widetilde{\Lc }} ( \bbe , \si ; \om ; \ep ) d( \bbe , \si ) = 0 \text{.} \non 
\end{align}
Fix $1 \le l \le \min \{ | \widetilde{\Lc } |, p \} $, $i_1 , \dots , i_l \in \widetilde{\Lc }$ with $i_1 < \dots < i_l$, and $1 \le k_1 < \dots < k_l \le p$. 
Let 
\begin{align}
A( \om ; \ep , \de ) &= \Big( \bigcap_{i \in \{ i_1 , \dots , i_l \} } \{ \bbe \in \mathbb{R} ^p | | a_i + b_i \om - {\x _i}^{\top } \bbe | \le \ep \om \} \Big) \non \\
&\quad \cap \Big( \bigcap_{i \in \widetilde{\Lc } \setminus \{ i_1 , \dots , i_l \} } \{ \bbe \in \mathbb{R} ^p | | a_i + b_i \om - {\x _i}^{\top } \bbe | > \ep \om \} \Big) \non \\
&\quad \cap \Big\{ \bigcap_{k \in \{ k_1 , \dots , k_l \} } \{ \bbe \in \mathbb{R} ^p | |( \e _{k}^{(p)} )^{\top } \bbe | \ge \de \om \} \Big\} \non \\
&\quad \cap \Big[ \Big( \bigcap_{j \in \widetilde{\Jc }} \{ \bbe \in \mathbb{R} ^p | | y_j - {\x _j}^{\top } \bbe | > \eta \} \Big) \non \\
&\quad \cup \bigcup_{1 \le q \le p - l} \bigcup_{\substack{j_1 , \dots , j_q \in \widetilde{\Jc } \\ j_1 < \dots < j_q}} \Big\{ \Big( \bigcap_{j \in \{ j_1 , \dots , j_q \} } \{ \bbe \in \mathbb{R} ^p | | y_j - {\x _j}^{\top } \bbe | \le \eta \} \Big) \non \\
&\quad \cap \Big( \bigcap_{j \in \widetilde{\Jc } \setminus \{ j_1 , \dots , j_q \} } \{ \bbe \in \mathbb{R} ^p | | y_j - {\x _j}^{\top } \bbe | > \eta \} \Big) \Big\} \Big] \text{.} \non 
\end{align}
As in the previous case, for some $\al < {\al }' \le \nu $ that is sufficiently close to $\al $, 
\begin{align}
&1( \bbe \in A( \om ; \ep , \de )) h_{\widetilde{\Kc } , \widetilde{\Lc }} ( \bbe , \si ; \om ) \non \\
&\le M_4 1( \bbe \in A( \om ; \ep , \de )) \pi ( \si ) \non \\
&\quad \times \Big\{ \prod_{k \in \{ 1, \dots , p \} \setminus \{ k_1 , \dots , k_l \} } {(| \be _k | / \si )^{\ka - 1} / \si \over (1 + | \be _k | / \si )^{\ka + \nu }} \Big\} \Big\{ \prod_{i \in \widetilde{\Kc }} {{\rm{N}} ( y_i | {\x _i}^{\top } \bbe , \si ^2 ) \over f_1 ( y_i )} \Big\} \Big( \prod_{i \in \{ i_1 , \dots , i_l \} } {\rm{N}} ( y_i | {\x _i}^{\top } \bbe , \si ^2 ) \Big) \non \\
&\quad \times \Big\{ \prod_{i \in \{ i_1 , \dots , i_l \} } {\om ^{\al } ( \log \om )^{1 + \ga } \over \om ^{{\al }'}} \Big\} \si ^{l {\al }'} [ \si ^{\al } \{ 1 + \log (1 + \si ) \} ^{1 + \ga } ]^{| \widetilde{\Lc } \setminus \{ i_1 , \dots , i_l \} |} \non 
\end{align}
for some $M_4 > 0$. 
Therefore, 
\begin{align}
&\int_{\mathbb{R} ^p \times (0, \infty )} 1\Big( \bbe \in \bigcap_{j \in \widetilde{\Jc }} \{ \bbet \in \mathbb{R} ^p | | y_j - {\x _j}^{\top } \bbet | > \eta \} \Big) 1( \bbe \in A( \om ; \ep , \de )) h_{\widetilde{\Kc } , \widetilde{\Lc }} ( \bbe , \si ; \om ) d( \bbe , \si ) \non \\
&\le M_5 \int_{\mathbb{R} ^p \times (0, \infty )} \Big( \pi ( \si ) \Big\{ \prod_{k \in \{ 1, \dots , p \} \setminus \{ k_1 , \dots , k_l \} } {(| \be _k | / \si )^{\ka - 1} / \si \over (1 + | \be _k | / \si )^{\ka + \nu }} \Big\} \Big( \prod_{i \in \{ i_1 , \dots , i_l \} } {\rm{N}} ( y_i | {\x _i}^{\top } \bbe , \si ^2 ) \Big) \non \\
&\quad \times \Big\{ \prod_{i \in \{ i_1 , \dots , i_l \} } {\om ^{\al } ( \log \om )^{1 + \ga } \over \om ^{{\al }'}} \Big\} \si ^{l {\al }'} [ \si ^{\al } \{ 1 + \log (1 + \si ) \} ^{1 + \ga } ]^{| \widetilde{\Lc } \setminus \{ i_1 , \dots , i_l \} |} \Big) d( \bbe , \si ) \to 0 \non 
\end{align}
as $\om \to \infty $. 
Now, suppose that $p \ge l + 1$ and fix $1 \le q \le p - l$ and $j_1 , \dots , j_q \in \widetilde{\Jc }$ with $j_1 < \dots < j_q$. 
Then if $\om > \eta / \de $, 
\begin{align}
&\int_{\mathbb{R} ^p \times (0, \infty )} \Big\{ 1 \Big( \bbe \in \Big( \bigcap_{j \in \{ j_1 , \dots , j_q \} } \{ \bbet \in \mathbb{R} ^p | | y_j - {\x _j}^{\top } \bbet | \le \eta \} \Big) \non \\
&\quad \cap \bigcap_{j \in \widetilde{\Jc } \setminus \{ j_1 , \dots , j_q \} } \{ \bbet \in \mathbb{R} ^p | | y_j - {\x _j}^{\top } \bbet | > \eta \} \Big) 1( \bbe \in A( \om ; \ep , \de )) h_{\widetilde{\Kc } , \widetilde{\Lc }} ( \bbe , \si ; \om ) \Big\} d( \bbe , \si ) \non \\
&\le M_4 \Big\{ \prod_{i \in \{ i_1 , \dots , i_l \} } {\om ^{\al } ( \log \om )^{1 + \ga } \over \om ^{{\al }'}} \Big\} \int_{\mathbb{R} ^p \times (0, \infty )} \Big\{ 1 \Big( \bbe \in \Big( \bigcap_{j \in \{ j_1 , \dots , j_q \} } \{ \bbet \in \mathbb{R} ^p | | y_j - {\x _j}^{\top } \bbet | \le \eta \} \Big) \non \\
&\quad \cap \bigcap_{j \in \widetilde{\Jc } \setminus \{ j_1 , \dots , j_q \} } \{ \bbet \in \mathbb{R} ^p | | y_j - {\x _j}^{\top } \bbet | > \eta \} \Big) 1( \bbe \in A( \om ; \ep , \de )) \pi ( \si ) \non \\
&\quad \times \Big\{ \prod_{k \in ( \{ 1, \dots , p \} \setminus \{ k_1 , \dots , k_l \} ) \cap \{ - j_1 , \dots , - j_q \} } {(| \be _k | / \si )^{\ka - 1} / \si \over (1 + | \be _k | / \si )^{\ka + \nu }} \Big\} \Big\{ \prod_{i \in \widetilde{\Kc } \cap \{ j_1 , \dots , j_q \} } {{\rm{N}} ( y_i | {\x _i}^{\top } \bbe , \si ^2 ) \over f_1 ( y_i )} \Big\} \non \\
&\quad \times \Big\{ \prod_{k \in ( \{ 1, \dots , p \} \setminus \{ k_1 , \dots , k_l \} ) \setminus \{ - j_1 , \dots , - j_q \} } {(| \be _k | / \si )^{\ka - 1} / \si \over (1 + | \be _k | / \si )^{\ka + \nu }} \Big\} \Big\{ \prod_{i \in \widetilde{\Kc } \setminus \{ j_1 , \dots , j_q \} } {{\rm{N}} ( y_i | {\x _i}^{\top } \bbe , \si ^2 ) \over f_1 ( y_i )} \Big\} \non \\
&\quad \times \Big( \prod_{i \in \{ i_1 , \dots , i_l \} } {\rm{N}} ( y_i | {\x _i}^{\top } \bbe , \si ^2 ) \Big) \si ^{l {\al }'} [ \si ^{\al } \{ 1 + \log (1 + \si ) \} ^{1 + \ga } ]^{| \widetilde{\Lc } \setminus \{ i_1 , \dots , i_l \} |} \Big\} d( \bbe , \si ) \non \\
&\le M_5 \Big\{ \prod_{i \in \{ i_1 , \dots , i_l \} } {\om ^{\al } ( \log \om )^{1 + \ga } \over \om ^{{\al }'}} \Big\} \int_{\mathbb{R} ^p \times (0, \infty )} \Big\{ 1 \Big( \bbe \in \Big( \bigcap_{j \in \{ j_1 , \dots , j_q \} } \{ \bbet \in \mathbb{R} ^p | | y_j - {\x _j}^{\top } \bbet | \le \eta \} \Big) \non \\
&\quad \cap \bigcap_{j \in \widetilde{\Jc } \setminus \{ j_1 , \dots , j_q \} } \{ \bbet \in \mathbb{R} ^p | | y_j - {\x _j}^{\top } \bbet | > \eta \} \Big) 1( \bbe \in A( \om ; \ep , \de )) \pi ( \si ) \si ^{l {\al }'} [ \si ^{\al } \{ 1 + \log (1 + \si ) \} ^{1 + \ga } ]^{| \widetilde{\Lc } \setminus \{ i_1 , \dots , i_l \} |} \non \\
&\quad \times \Big\{ \prod_{j \in ( \{ - 1, \dots , - p \} \setminus \{ - k_1 , \dots , - k_l \} ) \cap \{ j_1 , \dots , j_q \} } {(| \be _{- j} | / \si )^{\ka - 1} / \si \over (1 + | \be _{- j} | / \si )^{\ka + \nu }} \Big\} \Big\{ \prod_{i \in \widetilde{\Kc } \cap \{ j_1 , \dots , j_q \} } {{\rm{N}} ( y_i | {\x _i}^{\top } \bbe , \si ^2 ) \over f_1 ( y_i )} \Big\} \non \\
&\quad \times \Big\{ \prod_{j \in ( \{ - 1, \dots , - p \} \setminus \{ - k_1 , \dots , - k_l \} ) \setminus \{ j_1 , \dots , j_q \} } {(| \be _{- j} | / \si )^{\ka - 1} / \si \over (1 + | \be _{- j} | / \si )^{\ka + \nu }} \Big\} \Big( \prod_{i \in \{ i_1 , \dots , i_l \} } {\rm{N}} ( y_i | {\x _i}^{\top } \bbe , \si ^2 ) \Big) \Big\} d( \bbe , \si ) \non \\
&= M_5 \Big\{ \prod_{i \in \{ i_1 , \dots , i_l \} } {\om ^{\al } ( \log \om )^{1 + \ga } \over \om ^{{\al }'}} \Big\} \int_{\mathbb{R} ^p \times (0, \infty )} \Big\{ 1 \Big( \bbe \in \Big( \bigcap_{j \in \{ j_1 , \dots , j_q \} } \{ \bbet \in \mathbb{R} ^p | | y_j - {\x _j}^{\top } \bbet | \le \eta \} \Big) \non \\
&\quad \cap \bigcap_{j \in \widetilde{\Jc } \setminus \{ j_1 , \dots , j_q \} } \{ \bbet \in \mathbb{R} ^p | | y_j - {\x _j}^{\top } \bbet | > \eta \} \Big) 1( \bbe \in A( \om ; \ep , \de )) \pi ( \si ) \si ^{l {\al }'} [ \si ^{\al } \{ 1 + \log (1 + \si ) \} ^{1 + \ga } ]^{| \widetilde{\Lc } \setminus \{ i_1 , \dots , i_l \} |} \non \\
&\quad \times \Big\{ \prod_{j \in \{ - 1, \dots , - p \} \cap \{ j_1 , \dots , j_q \} } {(| \be _{- j} | / \si )^{\ka - 1} / \si \over (1 + | \be _{- j} | / \si )^{\ka + \nu }} \Big\} \Big\{ \prod_{i \in \widetilde{\Kc } \cap \{ j_1 , \dots , j_q \} } {{\rm{N}} ( y_i | {\x _i}^{\top } \bbe , \si ^2 ) \over f_1 ( y_i )} \Big\} \non \\
&\quad \times \Big\{ \prod_{j \in \{ - 1, \dots , - p \} \setminus ( \{ - k_1 , \dots , - k_l \} \cup \{ j_1 , \dots , j_q \} )} {(| \be _{- j} | / \si )^{\ka - 1} / \si \over (1 + | \be _{- j} | / \si )^{\ka + \nu }} \Big\} \Big( \prod_{i \in \{ i_1 , \dots , i_l \} } {\rm{N}} ( y_i | {\x _i}^{\top } \bbe , \si ^2 ) \Big) \Big\} d( \bbe , \si ) \text{,} \non 
\end{align}
where the equality follows since there is no point $\bbet = ( \bet _k )_{k = 1}^{p} \in \mathbb{R} ^p$ satisfying $| \bet _{- j} | \ge \de \om $ and $| y_j - {\x _j}^{\top } \bbet | \le \eta $ for some $j = - 1, \dots , - p$. 
The right-hand side converges to $0$ as $\om \to \infty $ regardless of whether $\{ - k_1 , \dots , - k_l \} \cap \{ j_1 , \dots , j_q \} = \emptyset $ or not. 
This completes the proof. 
\hfill$\Box$

\begin{remark}
Although we assume for simplicity that $f_0$ is the standard normal density, similar results about posterior robustness will be established for other choices of $f_0$ as well. 
The most important property of $f_0$ that is used throughout the above proof is that $f_0$ has thinner tails than $f_1$. 
\end{remark}

\section{Proof of Theorem 
2}

Here, we prove Theorem 
2.

\bigskip

\noindent
{\bf Proof of Theorem 
2.} \ \ As in the proof of Theorem 
1, we have 
\begin{align}
p( \bbe , \si | \y ) &= g( \bbe , \si ; \om ) / \int_{\mathbb{R} ^p \times (0, \infty )} g( \bbe , \si ; \om ) d( \bbe , \si ) \text{,} \non 
\end{align}
where 
\begin{align}
g( \bbe , \si ; \om ) %
&= \pi ( \bbe , \si ) \Big[ \prod_{i \in \Kc } \Big\{ {1 - s \over s} {{\rm{N}} ( y_i | {\x _i}^{\top } \bbe , \si ^2 ) \over f_1 ( y_i )} + 1 \Big\} \Big] \prod_{i \in \Lc } \Big\{ {1 - s \over s} {{\rm{N}} ( y_i | {\x _i}^{\top } \bbe , \si ^2 ) \over f_1 ( y_i )} + 1 \Big\} \text{,} \non 
\end{align}
and 
\begin{align}
\lim_{\om \to \infty } g( \bbe , \si ; \om ) = \pi ( \bbe , \si ) \prod_{i \in \Kc } \Big\{ {1 - s \over s} {{\rm{N}} ( y_i | {\x _i}^{\top } \bbe , \si ^2 ) \over f_1 ( y_i )} + 1 \Big\} < \infty \text{.} \non 
\end{align}
Now, if $\om $ is sufficiently large such that $| y_i | \le 2 | b_i | \om $ for all $i \in \Lc $, then 
\begin{align}
&\int_{\mathbb{R} ^p \times (0, \infty )} g( \bbe , \si ; \om ) d( \bbe , \si ) \non \\
&\ge \int_{\mathbb{R} ^p \times (0, \infty )} \pi ( \bbe , \si ) \prod_{i \in \Lc } \Big\{ {1 - s \over s} {{\rm{N}} ( y_i | {\x _i}^{\top } \bbe , \si ^2 ) \over f_1 ( y_i )} \Big\} d( \bbe , \si ) \non \\
&\ge {1 \over M_1} \om ^{| \Lc | (1 + \al )} \int_{\mathbb{R} ^p \times (0, \infty )} \pi ( \si ) {1 \over \si ^{| \Lc |}} \pi ( \bbe | \si ) \Big[ \prod_{i \in \Lc } \exp \Big\{ - {2 (| y_i |^2 + | {\x _i}^{\top } \bbe |^2 ) \over 2 \si ^2} \Big\} \Big] d( \bbe , \si ) \non \\
&\ge {1 \over M_1} \om ^{| \Lc | (1 + \al )} \int_{\{ ( \bbet , \sit ) \in \mathbb{R} ^p \times (0, \infty ) | \| \bbet \| \le \si \ge \om \} } \pi ( \si ) {1 \over \si ^{| \Lc |}} {h( \bbe / \si ) \over \si ^p} \Big\{ \prod_{i \in \Lc } \exp \Big( - {4 | b_i |^2 \om ^2 \over \si ^2} - \| \x _i \| ^2 \Big) \Big\} d( \bbe , \si ) \non \\
&\ge {1 \over M_2} \om ^{| \Lc | (1 + \al )} \int_{\om }^{\infty } {1 \over \si ^{| \Lc | \al + 1 - \rho }} {1 \over \si ^{| \Lc |}} \exp \Big( - \sum_{i \in \Lc } {4 | b_i |^2 \om ^2 \over \si ^2} \Big) d\si \text{.} \non 
\end{align}
Therefore, by making the change of variables $\si = \om s$, we obtain 
\begin{align}
&\lim_{\om \to \infty } \int_{\mathbb{R} ^p \times (0, \infty )} g( \bbe , \si ; \om ) d( \bbe , \si ) \ge \lim_{\om \to \infty } {\om ^{\rho } \over M_2} \int_{1}^{\infty } {1 \over s^{| \Lc | (1 + \al ) + 1 - \rho }} \exp \Big( - \sum_{i \in \Lc } {4 | b_i |^2 \over s^2} \Big) d\si = \infty \text{.} \non 
\end{align}
This completes the proof. 
\hfill$\Box$

\end{document}